\documentclass[journal]{IEEEtran}
\usepackage{graphicx}
\usepackage{pzccal}
\usepackage{mathrsfs}

\usepackage{float}

\usepackage{bbding}
\usepackage{soul}
\usepackage{xcolor}

\usepackage{float}

\usepackage{cite}
\ifCLASSINFOpdf
\else
\fi
\usepackage[cmex10]{amsmath}
\usepackage{algorithmic}
\usepackage{array}
\usepackage{url}
\begin{document}

\title{A Complete Model for Modular Simulation of Data Centre Power Load}

%\author{\IEEEauthorblockN{R. Rahmani* and I. Moser}
%\IEEEauthorblockA{Faculty of Science, Engineering and Technology \\
%Swinburne University of Technology \\ Hawthorn, VIC 3122, Australia \\
%Emails: *rrahmani@swin.edu.au \\
%imoser@swin.edu.au}
%\and
%\IEEEauthorblockN{M. Seyedmahmoudian}
%\IEEEauthorblockA{School of Engineering, Deakin University \\ Waurn Ponds, Victoria 3216, Australia\\
%Email: mseyedma@deakin.edu.au}}



\author{R.~Rahmani,~\IEEEmembership{Member,~IEEE,}
	I.~Moser,~\IEEEmembership{Member,~IEEE,}
	and~M.~Seyedmahmoudian,~\IEEEmembership{Member,~IEEE}% <-this % stops a space
	\thanks{R. Rahmani and I. Moser are with the Faculty of Science, Engineering and Technology, Swinburne University of Technology, Hawthorn, VIC 3122, Australia e-mails: \{rrahmani,imoser\}@swin.edu.au.}% <-this % stops a space
	\thanks{M. Seyedmahmoudian is with School of Engineering, Deakin University, Waurn Ponds, VIC 3216, Australia.}% <-this % stops a space
	\thanks{Manuscript received April 19, 2017; revised August 26, 2017.}}

\markboth{Journal of IEEE Transactions on Automation Science and Engineering,~Vol.~14, No.~8, August~2017}%
{Shell \MakeLowercase{\textit{et al.}}: Bare Demo of IEEEtran.cls for IEEE Journals}


\maketitle

\begin{abstract}
%\boldmath
Data centres are very fast growing structures with significant contribution to the world's energy consumption. Reducing the energy consumption of data centres is easier when the components that comprise a data centre and their respective energy consumption are known.
A complete model for a modular design of a data centre and a technique for simulating each module's energy consumption are presented. 
Detailed power consumption modelling for each component as well as their interactions are the merits of this model. Unlike existing research, the present modular simulation model can take different design structures of data centres into account and provide us with hourly power consumption profiles for each component. The impacts of environmental parameters such as temperature and humidity are also investigated and incorporated into the model. The flexibility, scalability, comprehensiveness and modularity of this model provides researchers and designers with a powerful tool for energy analysis, management and planning of data centres with different designs and locations.
\vspace{0.1in}

\textit{Note to Practitioners - }The main goal of this paper is to propose a modular approach for designing a typical data centre. The total power consumption profile of any data centre can be calculated based on considering its components and environmental data. By including and removing components, we can determine the consumption profile of a data centre at the design stage.
This model is intended as a tool for designers and planners, that defines the impact of each parameter on the power consumption of each component. Therefore, the effective parameters of each component are determined and placed into context to enhance the flexibility, scalability, comprehensiveness and modularity of this model towards a powerful tool for the practitioners in this field.
%\blindtext[1]
\end{abstract}


\begin{IEEEkeywords}
Data Centre, power load modelling, server utilisation, modular simulation, mathematical modelling. 
\end{IEEEkeywords}

% \fi
\IEEEpeerreviewmaketitle

\section{Introduction}
%\blindtext

The growth of demand for information and communication technology (ICT) services and their associated concerns such as data security and privacy has increased the need for designing and constructing sophisticated data centres. In 2014, data centres consumed 2\% of the total energy consumption of USA, a staggering 70 billion kWh. This was equal to the energy demand for 6.4 million average American households. The energy demand for data centres doubled during the last five years \cite{arnone2015smart,datacenterknowledge}. In Australia, these gigantic structures consumed 3.9\% of the national electricity consumption in 2013 while it is projected that internet traffic will grow 36\% per year \cite{energyratingaustralia}.

The rapid worldwide growth of data centres and the resultant carbon emissions have spurred governmental and organisational bodies into setting energy efficiency criteria for data centre infrastructure design \cite{shuja2014data}. In the light of this fact, researchers have paid much attention to the problem of enhancing or optimising the performance of different aspects and sections in data centres including data scheduling and cooling infrastructure \cite{pelley2009understanding,fang2016qos}. Notwithstanding the volume of research dedicated to this field, most of the literature focuses on modelling and analysing subsystems of a data centre as independent entities without taking into consideration their interconnections and the influence they have on the energy consumption of other parts of the data centre \cite{ebrahimi2014review,oro2015energy}.

This paper represents a modular simulation model for a data centre with $\mathcal{N}$ servers. The present model consists of detailed power consumption modelling for each component as well as the interactions between all components. Unlike existing research in the literature, the present modular simulation model can take different design structures of data centres into account and provide us with hourly power consumption profiles for each component. The present power consumption model is independent of the main source of electricity of the data centre which can be supplied by the main grid or by renewable energies through microgrids. The flexibility, scalability, comprehensiveness and modularity of this model provides the researchers and designers with a powerful tool for energy analysis, management and planning of data centres with different designs and locations.

\vspace{-0.1in}
\section{Components of a Data Centre}
In general, data centres are huge infrastructures which can occupy hundreds of thousands of square feet, e. g., Facebook's Prineville data centre in Oregon consists of three buildings which nearly cover 800,000 square feet of land. This massive construction is meant to provide services like processing, networking, storage, management and distribution of data and information to public or organisations. The main source of electricity for a data centre is usually the grid connection which is provided by utility companies, although there are some exceptions like Apple's data centres which claim to use 100\% renewable energy. Although the components of a data centre depend on its design and structure, the following components and systems are generally present in a typical data centre.

\vspace{-0.1in}
\subsection{Main Components and Equipment}

\subsubsection{Server farm} This is considered the main part of a data centre as it provides the ICT service of a data centre. and its size is the most important parameter in determining the energy requirement of a data centre as a whole. A server farm consists of racks, with each rack containing tens of servers. Each server is usually assumed to be 1 U in height (1 U equals 44.45 mm) and a cluster is built by stacking them. The server farm is the major energy consumption unit in any data centre.

\subsubsection{Power supply equipment} The electricity of a data centre is often provided by a high voltage transformer (in the range of a few tens of kV) which feeds an uninterruptible power supply (UPS) unit. The UPS can provide the data centre with power only for the short periods while backup generators start up during utility supply failure. The power is supplied to the IT equipment using power distribution units (PDUs) which regulate the voltages below 1 kV, depending on the server rack's demand.

\subsubsection{Backup generation} Due to the possible failure of supply from utility companies, each data centre is equipped with a backup generation system which is mostly based on diesel generators. The backup energy supply must be able to support at least the vital IT loads of the data centre for a reasonable period of time.

\subsubsection{Energy storage}
This is another section considered in the design of any data centre. Its most important benefit to the control and management system is having almost zero start-up time. As a power supply, the energy storage can instantaneously provide electricity which makes it valuable as a backup system without start-up delay. In data centre design, usually battery systems are used as energy storage.

\subsubsection{Heat and air flow system}
The heat generated by the servers must be removed before the ambient temperature of the server increases to a critical value, as elevated temperatures around servers can damage them. Two common types of heat and air flow systems are used in data centres which are the computer room air handler (CRAH) and computer room air conditioning (CRAC) units. The main difference is that CRAC units include their own condensers for reducing the air temperature while CRAH units need an external cooling system to provide them with chilled water. The main task of a CRAC or CRAH unit is to ensure that the ambient temperature of the servers is within 20 to 25 $^\circ$C, with an allowable range spanning 15 to 32 $^\circ$C as recommended by the American Society of Heating, Refrigerating and Air-conditioning Engineers (ASHRAE) \cite{handbook2001fundamentals}. 

The air flow in a computer room is usually from bottom to top and the server racks are usually positioned so that they form cold and hot air aisles. The cold air is thrust upward to the cold aisles, where servers swallow it and return the hot air to the hot air aisles. The hot air is gathered at the hot air plenum and returned to the air chase for being cooled. 

%Fig. \ref{airFlow} WILL BE REPLACED

%\begin{figure}[ht]
%	\centering
%	\includegraphics[width=2.5in]{airFlow.jpg}%
	%\subfloat[Case II]{\includegraphics[width=2.5in]{subfigcase2}%
	%\label{fig_second_case}}}
%	\caption{Computer room air flow of a typical data centre performed by a CRAH or CRAC unit.}
%	\label{airFlow}
%\end{figure}

\subsubsection{Cooling system}
The cooling system of data centres is commonly based on a water chiller plant in which a vapour-compression or absorption cycle is used to reduce the temperature of the cooling fluid. There are different techniques for transferring heat between the data centre and the cooler using transport fluid in data centres. Evans \cite{evans2012different} explains these in detail. After the server farm, the cooling system is usually the second most significant power consumer in a data centre. The impact of climate parameters such as ambient temperature is considerable on the power consumption of a chiller plant unit. There are exceptions in utilising cooling systems in data centres. Google's Data Centre in Hamina, Finland is located beside a natural river and the water flow from the river is used without applying any temperature control system. Another example is Facebook's Data Centre in Lule{\aa}, Sweden which is claimed to be the most efficient and sustainable to date, as naturally cold air is used to flow to the computer rooms.

\subsubsection{Control and security system}
Although the power consumption of this unit is a few percent of the whole, its contribution into the safety and security of the data stored is significant and the model of its energy consumption is included for the sake of completeness. This unit is responsible for data and power flow control and management, and also making critical decisions about data at the time of emergency.

%\subsection{Data Flow}

%\begin{figure}[ht]
%	\includegraphics[width=2in]{dataFlow.png}%
	%\subfloat[Case II]{\includegraphics[width=2.5in]{subfigcase2}%
	%\label{fig_second_case}}}
%	\caption{Data Flow}
%	\label{dataFlow}
%\end{figure}

\vspace{-0.1in}
\subsection{Interconnection of components}

\begin{figure}[ht]
\centering
\includegraphics[width=3.5in]{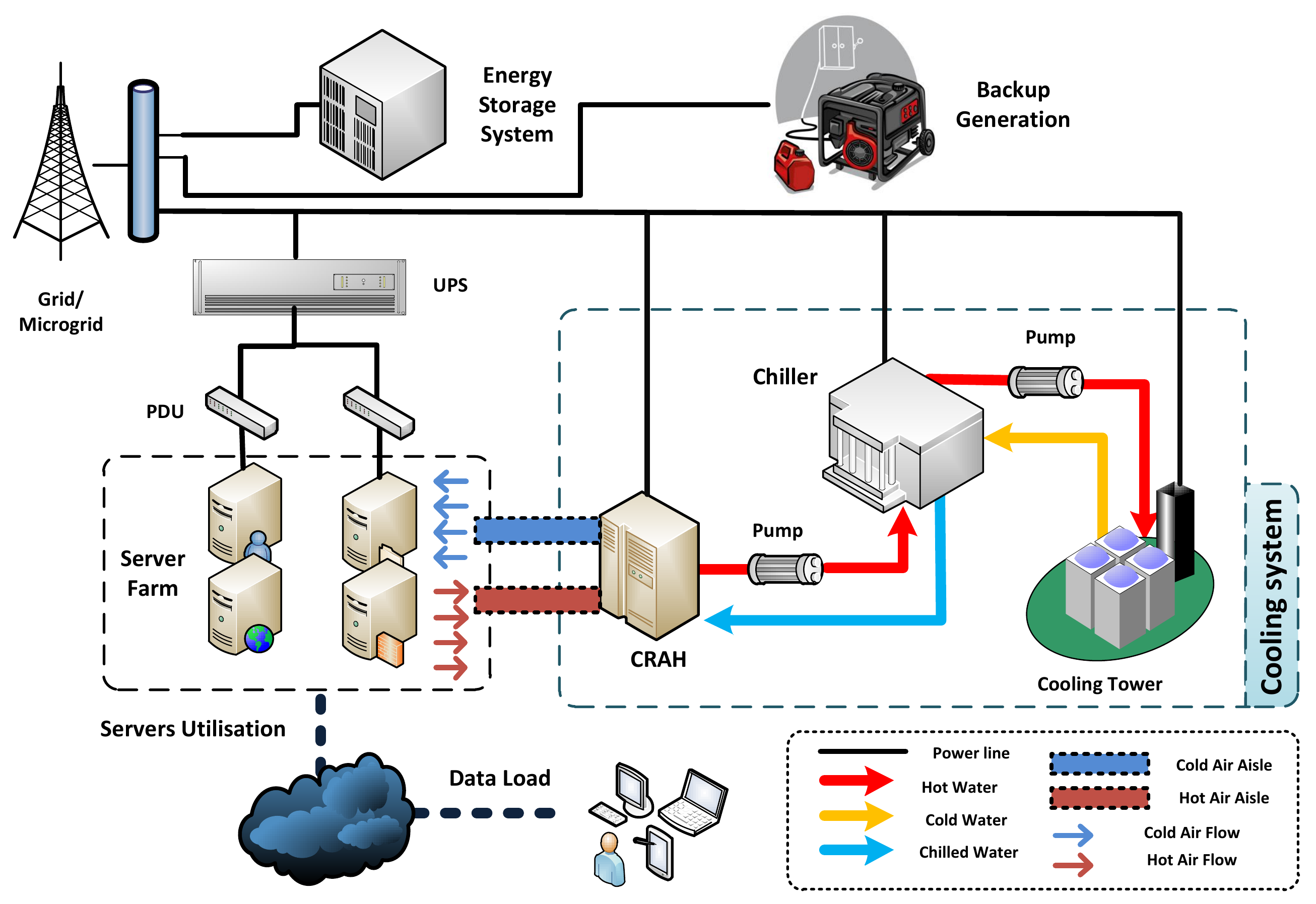}%
%\subfloat[Case II]{\includegraphics[width=2.5in]{subfigcase2}%
%\label{fig_second_case}}}
\caption{The interconnection of components in a typical data centre.}
\label{powerFlow}
\end{figure}

The data, air and power flow, as well as the interconnection of all components in a typical data centre,  are shown in Fig. \ref{powerFlow}. The cooling system is assumed to be based on chilled water. No matter what architecture a data centre might have, it has only two types of connections with the outer world - electrical power and internet connections. The data load determines server utilisation and the total power demand changes based on the data load. How do the interconnections of the components in a data centre affect the total power consumption? To answer these question, we need to develop detailed models for all components and their interconnections.

\vspace{-0.1in}
\section{Modelling and Simulation of the Components}
\label{modelSection}

The main goal of this paper is to propose a modular approach for designing a typical data centre. The total power consumption profile of any data centre can be calculated based on considering its components and environmental data. By including and removing components, we can determine the consumption profile of a data centre at the design stage.
This model is intended as a tool for designers and planners, that defines the impact of each parameter on the power consumption of each component. Therefore, the effective parameters of each component are determined and placed into context in this section.  

\vspace{-0.1in}
\subsection{Information Technology Equipment}
\label{serverFarmSection}
The power consumption of this subcomponent of the data centre mainly depends on the number of servers ($\mathcal{N}$) and total utilisation of all servers ($\mathcal{U}$). The effects of task scheduling and load balancing therefore need to be reflected in those main two parameters. 

The utilisation of the $i^{th}$ server ($u_{i}$) is considered unitless and varies between 0 and 1, with 1 representing the full utilisation of the respective server. Extensive explanations and analyses on determining the unit for server utilisation based on computing speed (in MIPS) and MSRP resource allocation are available in the literature \cite{chase2001managing,giri2010increasing,buyya2010energy}. The parameters affecting a server's utilisation are basically its workload and consolidation mechanisms which are studied in several publications \cite{nathuji2008coolit,vogels2008beyond,verma2009server}.

If the server cluster is assumed to be homogeneous, the degree of utilisation can be expressed as:
\begin{equation}
\label{utilTotal}
\mathcal{U} = \frac{1}{\mathcal{N}}\sum_{i=1}^{\mathcal{N}}{u_i}
\end{equation}
where $\mathcal{N}$ represents the number of servers.

The server power consumption profile has been studied by several researchers \cite{rivoire2008comparison,barroso2013datacenter,meisner2009powernap} who state that a server consumes almost half of the rated power in idle mode with an approximately linear increase with utilisation:
\begin{equation}
\label{serverPower}
	p_i = p_i^{idle}+(p_i^{peak}-p_i^{idle})\times u_i
\end{equation}

\begin{figure}[ht]
	\centering
	\includegraphics[width=3in]{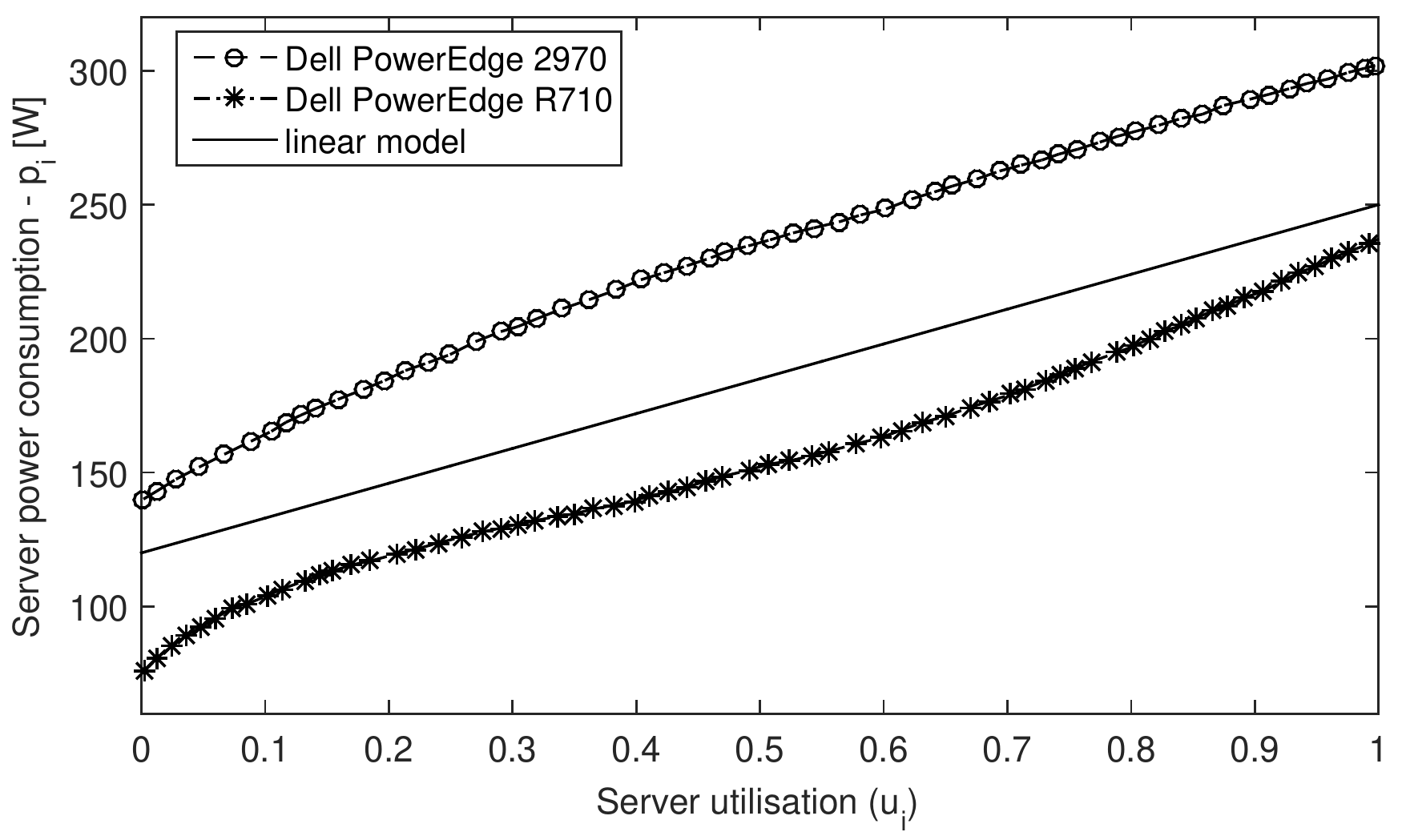}%
	%\subfloat[Case II]{\includegraphics[width=2.5in]{subfigcase2}%
	%\label{fig_second_case}}}
	\caption{Server power consumption versus utilisation.}
	\label{serverCons}
\end{figure}

Fig. \ref{serverCons} shows the linear approximation along with published SpecPower benchmark results for two Dell systems \cite{lange2009identifying}. In general, the total power consumption of a server farm ($\mathcal{P}_{sf}$) can be obtained from (\ref{PserverFarm}) where $p_i$ represents the power consumption of the $i^{th}$ server. However, it is not practical to measure the actual power consumption of each server rack continuously. Therefore, the impact of task consolidation ($\mathcal{L}$) for the running servers is modelled as in (\ref{consol}).

\begin{figure}[ht]
	\centering
	\includegraphics[width=3in]{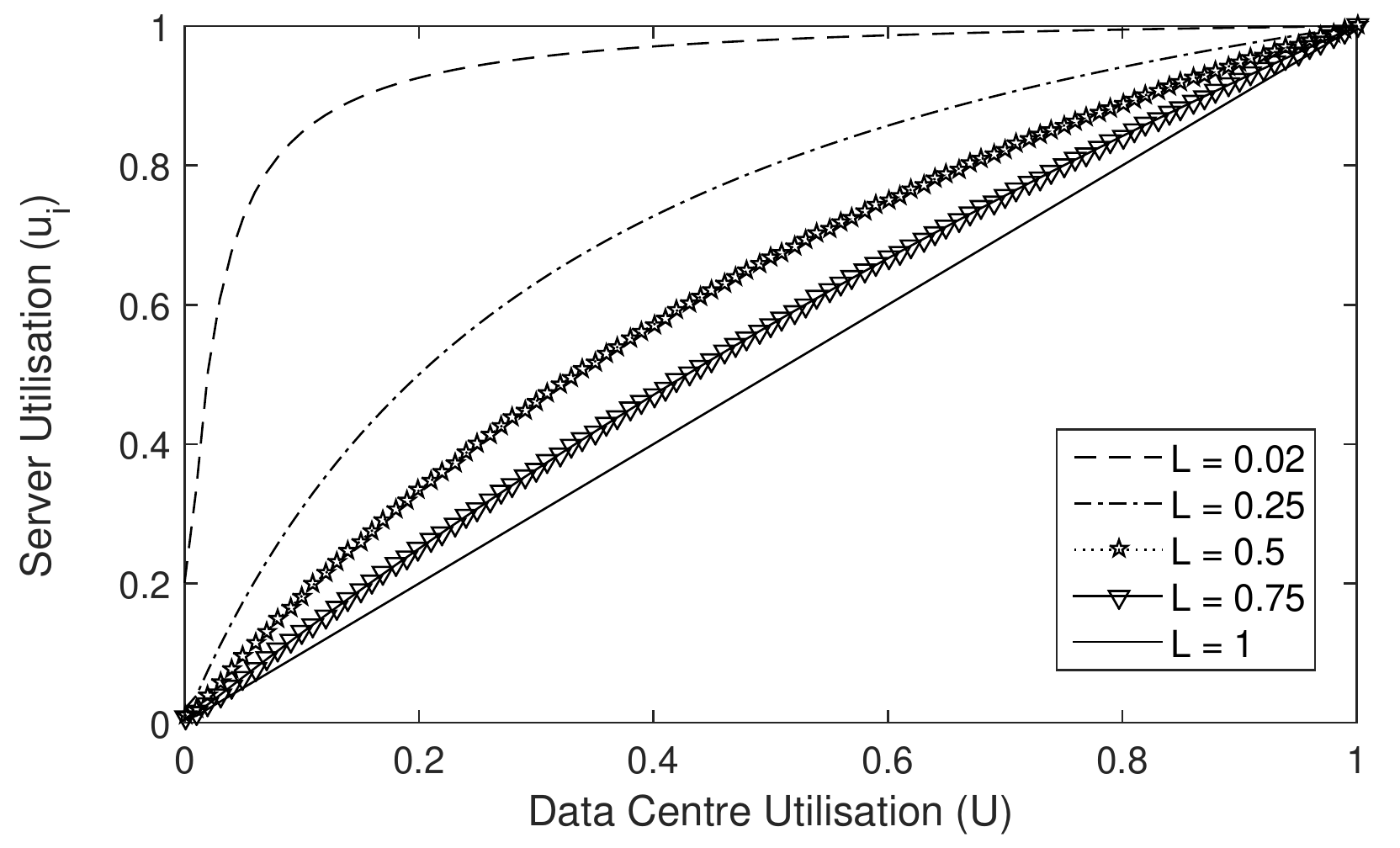}%
	%\subfloat[Case II]{\includegraphics[width=2.5in]{subfigcase2}%
	%\label{fig_second_case}}}
	\caption{Impact of consolidation on the relation between $u_s$ and $\mathcal{U}$.}
	\label{consolU}
\end{figure}

\vspace{-.15in}
\begin{equation}
\label{PserverFarm}
\mathcal{P}_{sf} = \sum_{i=1}^{\mathcal{N}}{p_i}
\end{equation}
\vspace{-0.07in}
\begin{equation}
\label{consol}
u_i = \frac{\mathcal{U}}{\mathcal{L}\times (1-\mathcal{U}) + \mathcal{U}}  
\end{equation}
where the perfect consolidation ($\mathcal{L}=0$) happens whenever the workload is packed onto the minimum number of running servers and perfect load balancing ($\mathcal{L}=1$) occurs when the workload is distributed uniformly to all servers and so $\mathcal{U}=u_i$ \cite{verma2009server,beloglazov2012optimal}. In reality, we have neither perfect consolidation, nor perfect load balancing; however, the model enables us to consider the consolidation effects on the power consumption of the server farm. The impact of consolidation on the relation between $u_i$ and $\mathcal{U}$ is depicted in Fig. \ref{consolU}.

Overall, the server farm's power consumption can be described as a function of data centre utilisation and consolidation ($\mathcal{P}_{sf}= f_{sf}(\mathcal{U},\mathcal{L})$).

\vspace{-.1in}
\subsection{Power supply equipment}
\label{powerSupplySection}
\begin{figure}[ht]
	\centering
	\includegraphics[width=3in]{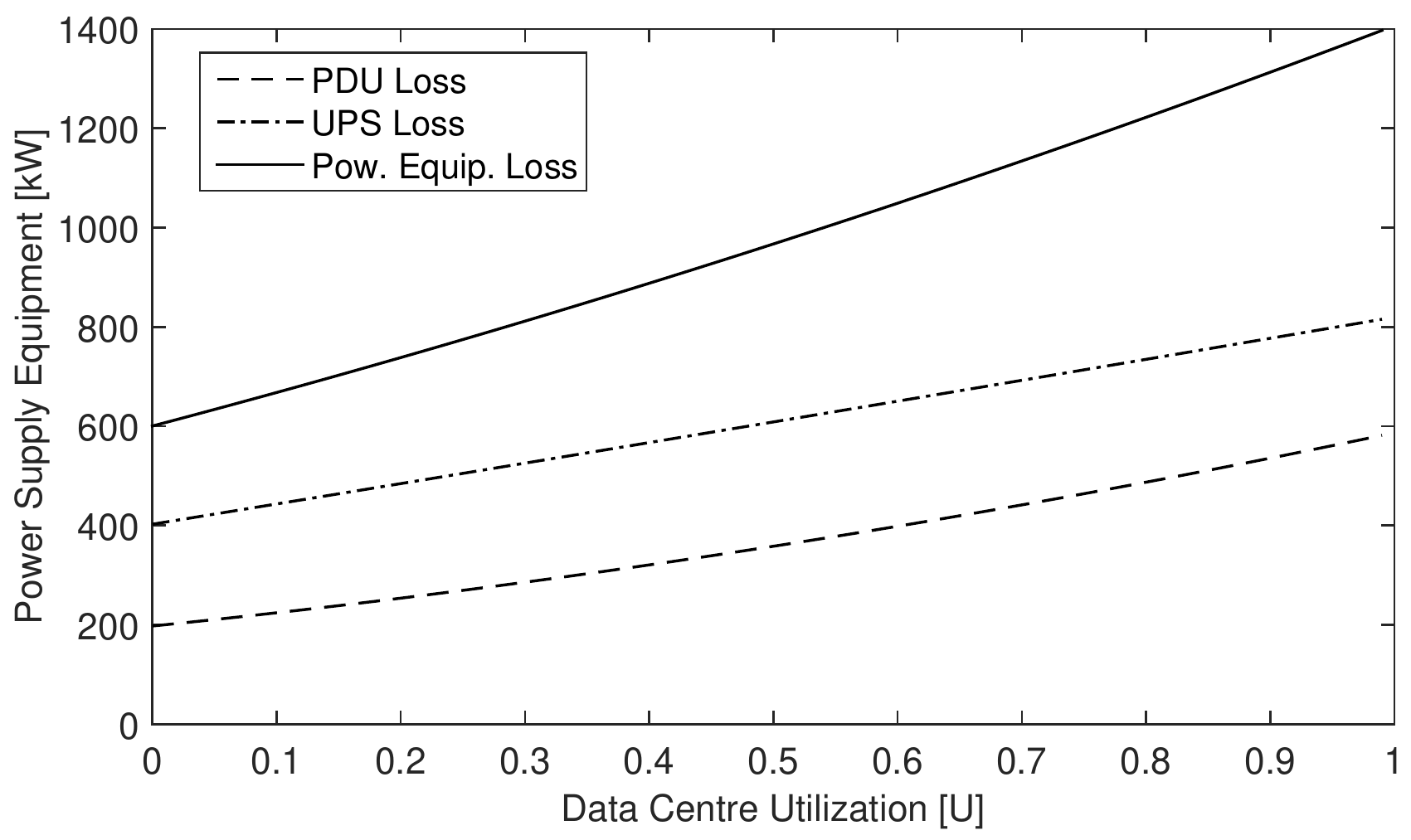}%
	%\subfloat[Case II]{\includegraphics[width=2.5in]{subfigcase2}%
	%\label{fig_second_case}}}
	\caption{Power loss of power supply equipment versus data centre utilisation.}
	\label{equipLoss}
\end{figure}

The amount of power loss in power supply equipment of a data centre can be as much as 15\% of the total power consumption at peak. Two main components involved in this power loss are the UPS and the PDUs. As can be observed in Fig. \ref{powerFlow}, the UPS supplies power to the PDUs which dispatch the electricity to the server racks. Eqs. (\ref{PDU}) and (\ref{UPS}) show the power loss of each of the PDU and the UPS respectively \cite{rasmussen2006electrical}.
\begin{equation}
\label{PDU}
	\mathcal{P}_{PDU} = 	\mathcal{P}_{PDU}^{idle} + \lambda_{PDU} (\sum_{\mathcal{N}_{run}} p_i )^2
\end{equation}
\begin{equation}
\label{UPS}
	\mathcal{P}_{UPS} = \mathcal{P}_{UPS}^{idle} + \lambda_{UPS}  \sum_{PDUs} \mathcal{P}_{PDU}
\end{equation}
were $\lambda_{PDU}$ and $\lambda_{UPS}$ represent the loss coefficients for PDU and UPS units respectively. The power consumption of each PDU may vary depending on the power consumption of the servers connected to it; however, due its low power loss rate (a few percent of the power consumption of the data centre), assuming perfect load balancing for its power calculation is acceptable. Therefore, the summation of all $\mathcal{P}_{PDU}$ values will result in the total power consumed by the PDU units. Both PDU and UPS units consume some power in idle mode which is reflected in the $\mathcal{P}^{idle}$ factor. Therefore, as in the case of the server farm, the power loss of the power equipment can be described as a function of the utilisation and consolidation of the servers.
 Fig. \ref{equipLoss} shows the power loss of the modelled power supply equipment versus data centre utilisation for a 10 MW data centre.

\vspace{-0.1in}
\subsection{Chiller plant}
\label{chillerSection}
Chiller plants in data centres perform temperature reduction for water or glycol as coolants. Although several types of chillers exist, such as vapour-compression and absorption chillers, a typical chiller plant consists of pumps, compressor, cooling tower and blower, evaporator and condenser. The utilisation of absorption chillers is more common in data centres which use a heat source for driving the refrigeration cycle instead of a condenser. The main parameter affecting the power consumption of a chiller plant is the amount of heat to be removed from the coolant which is related to the difference between the computer room's temperature and the required temperature value of the cold air aisle. This parameter is reflected in the temperature difference between the cold water being pumped to the computer room and hot fluid leaving the CRAH unit. Based on the thermodynamic principle, the general heat transfer equation for a CRAH unit is shown in (\ref{Qtrans}).
\begin{equation}
\label{Qtrans}
	\mathcal{Q}=\mathcal{K}\dot{m}C_{p}^{air}(T_h-T_c)
\end{equation}
where $\mathcal{Q}$ is the amount of heat which needs to be removed, $\dot{m}$ is the total air flowing through the servers, $C_p^{air}$ is the specific heat transfer capacity of air, $T_h$ and $T_c$ denote hot and cold air temperatures respectively. $\mathcal{K}$ is the containment index which represents the fraction of air ingested by the servers from the cold aisle. Ideally $\mathcal{K}=1$ denotes perfect containment or no air recirculation \cite{tozer2009air}.

One approach is to calculate the power consumption of the chiller plant based on the physical properties and amount of required heat to be removed from the coolant liquid ($\mathcal{Q}$), a.k.a physics-based model as has been presented in the literature \cite{zhangenergy,lu2015optimization,chen2015performance}. On the other hand, using an empirical model comprising of regression and curve fitting techniques is also of interest and sufficiently accurate \cite{browne1998challenges,lazrak2016development}. It can be said that the data centre utilisation is the main factor that causes servers to produce more or less heat which affects the computer room air temperature. Therefore, we choose the Department of Energy's DOE2 chiller model \cite{york1980doe} to derive a power consumption model as a function of $\mathcal{U}$. We do not explain the details of the DOE2 model as it is well documented and further investigation on chiller operation is not in the scope of this research work. The chiller's power consumption quadratically increases with the amount of heat to be removed and thus with the data centre utilisation. Sawyer \cite{sawyer2004calculating} estimated that the size of chiller plant has to be $0.7\times \mathcal{P}_{sf}^{max}$ in order to provide sufficient cooling; therefore, the chiller plant power consumption model is derived as shown in (\ref{chillerModel}).
\begin{equation}
\label{chillerModel}
	\mathcal{P}_{chiller} = 0.7\times \mathcal{P}_{sf}^{max} (\alpha \cdot \: \mathcal{U}^2+\beta \cdot \: \mathcal{U}+ \gamma)
\end{equation}
where $\alpha=0.32$, $\beta=0.11$ and $\gamma=0.63$ are obtained by performing curve fitting for several samples of real data centres in which the cooled water temperature is kept steady at 7.22$^\circ$C. The power consumption of the chiller plant is highly dependent on the chilled water temperature. As an example, if the supplied chilled water temperature is increased to 12.78$^\circ$C, the power consumption can drop by about 15\% at the peak load \cite{pelley2009understanding}. 

Water pumps are the equipment used in any water based cooling system like chillers. On average, the power consumption of the water pumps in a data centre is about one sixth of the total power consumption of the whole data centre \cite{sawyer2004calculating}.

\vspace{-.1in}
\subsection{CRAH and CRAC units}
\label{CRAHSection}
The main difference between the two types is that a CRAC unit (also known as CRAC direct-expansion (DX) system) includes a condenser while a CRAH unit mainly circulates the air to the chilled water supplied by the chiller plant. A CRAC unit does not require a chiller plant. Only a cooling tower is required for transferring the heat to the outside; therefore, its power consumption is higher than that of a CRAH unit. It is estimated that the total power consumption of a cooling system of a data centre which uses a CRAC system is higher compared with the ones that use CRAH and a chiller plant \cite{sawyer2004calculating}. Extensive experimental research has been performed by Itoh \textit{et al.} \cite{itoh2010power} who investigated the impacts of several parameters on the power consumption of a CRAC unit, such as the arrangement of the server racks and air temperature around them. They observed that the power consumption required for heat cycling is linearly proportional to the volume of the air flow ($\mathcal{f}$) and the required air flow in a server room can be obtained by $\mathcal{f} = \mathcal{f}^{max} \times \mathcal{U}$, where $\mathcal{f}^{max}$ is the maximum standard air flow (14000 CMH for a 7.5 kW CRAH unit). (\ref{heatModel}) shows the electrical power required for heat transfer to be consumed by an air handling system for a server farm with the maximum servers load of $\mathcal{P_{sf}^{max}}$ kW.
\begin{equation}
	\label{heatModel}
	\mathcal{P}_{heat} = 1.33\times 10^{-5} \times \frac{\mathcal{P_{sf}^{max}}}{\eta_{heat}}  \times \mathcal{f}
\end{equation}
were $\eta_{heat}$ denotes the efficiency of the removal of heat from the system. The total power required for a CRAH unit can be calculated using (\ref{crahModel}) where $\mathcal{P}_{CRAH}^{idle}$ can be considered as about 7 to 10 percent of $\mathcal{P}_{sf}^{max}$ \cite{itoh2010power}.
\begin{equation}
\label{crahModel}
\mathcal{P}_{CRAH} = \mathcal{P}_{CRAH}^{idle}+	\mathcal{P}_{heat}
\end{equation}
In order to obtain the power consumption of a CRAC system, the power consumption model of a condenser system has to be taken into account, in addition to the heat removing system. Based on the experimental research performed by Hajidavalloo and Eghtedari \cite{hajidavalloo2010performance}, the total power consumption of a CRAC by taking the air-cooled condenser into account can be written as (\ref{cracModel}).
\begin{equation}
	\label{cracModel}
	\mathcal{P}_{CRAC} = \mathcal{P}_{CRAC}^{idle}+	((1+COP)\times \mathcal{P}_{heat})
\end{equation} 
were $COP$ is the coefficient of performance for the condenser which is usually between 3 and 6. The lower COP denotes higher efficiency for the condenser of the CRAC unit. $\mathcal{P}_{CRAC}^{idle}$ is the power consumption of the CRAC system in idle mode which is normally between 10 to 30 percent of $\mathcal{P}_{sf}^{max}$.

\vspace{-.1in}
\subsection{Miscellaneous power consumption}
This consumption type includes all the lighting, security devices, control and monitoring systems, and networking facilities power consumption. The power consumption of this category is independent of server utilisation or environmental parameters, but it is understood that the only parameter that affects this consumption category is the size of data centre. The total power consumption of this category is considered to be 6\% of peak demand of the data centre \cite{pelley2009understanding}.

\vspace{-0.1in}

\section{Impacts of Environmental Parameters}
\label{environImpactSection}

According to the laws of thermodynamics, the parameters of the environment of data centre, such as ambient temperature, have impacts on the energy required for transferring the heat generated by the server farms from the computer room to the outside. No matter what cooling system is used in a data centre, its last element in the heat flow cycle is the cooling tower which is directly in contact with the ambient air. The ambient temperature and humidity are the two important parameters affecting the efficiency and power consumption of cooling systems in data centres. Considering that the computer room temperature has to be kept almost constant, the outdoor temperature plays the key role in the cooling load. The energy required for deposing the heat in a hotter environment is higher \cite{herold2016absorption}. Table \ref{EER} shows the energy efficiency ration (EER) of a chiller plant as a function of the ambient temperature \cite{salom2014cost}. In order to consider the effect of the ambient temperature on the power consumption of the cooling system, the power consumption has to be divided by EER.

\begin{table}[t]
\centering
\caption{Energy efficiency ratio (EER) of a chiller plant as a function of outdoor ambient temperature \cite{salom2014cost}.}
\label{EER}
\resizebox{\linewidth}{!}{
\renewcommand{\arraystretch}{1.5}
\begin{tabular}{l|ccccccccc} 
	Ambient Temperature ($^\circ$C)  & 41 & 35 & 30 & 25 & 20 & 15 & 10 & 5 & 0 \\ \hline
	EER & 2.66 &3.12 & 3.52 & 3.93 & 4.34 & 4.74 & 5.13 & 5.49 & 5.82 
\end{tabular}}
\end{table}

The humidity of the air has an impact on the power consumption of the cooling system by affecting the thermal conductivity of the air. As investigated by Tsilingiris \cite{tsilingiris2008thermophysical}, the change in the thermal conductivity of air due to humidity becomes significant after about 50$^\circ$C. At a constant temperature of 30$^\circ$C, the thermal conductivity only increases by 1.5\% if the relative humidity (RH) is changed from 100 to 0\%. This change is 1.8\% for 40$^\circ$C, and 3.4\% for 50$^\circ$C. By increasing the air temperature to 60$^\circ$C, the thermal conductivity may vary up to 5.5\% and the highest change of 26.2\% is experienced at 100$^\circ$C  when the RH is changed from 100\% to 0\%. Since the ambient temperature normally does not surpass 50$^\circ$C, the impact of ambient air humidity can be neglected in the power consumption modelling of a data centre.

\vspace{-0.1in}
\section{Modelling a Data Centre Power Consumption}

Generally, the design structure of a data centre determines its components, and the total power load profile of any data centre depends on the type and consumption of its components. For instance, if the cooling system of a data centre is chosen to be CRAC, the total power model and profile of the cooling system is different from a design with CRAH and chiller system. Also, if a data centre is located in a cold area and uses natural cold air to transfer heat, the cooling system may only consist of some fans for air ventilation. The modelling presented in this paper makes it possible to easily obtain the total load profile of different data centre designs by considering or eliminating components in the model.

\begin{figure}[ht]
	\centering
	\includegraphics[width=3in]{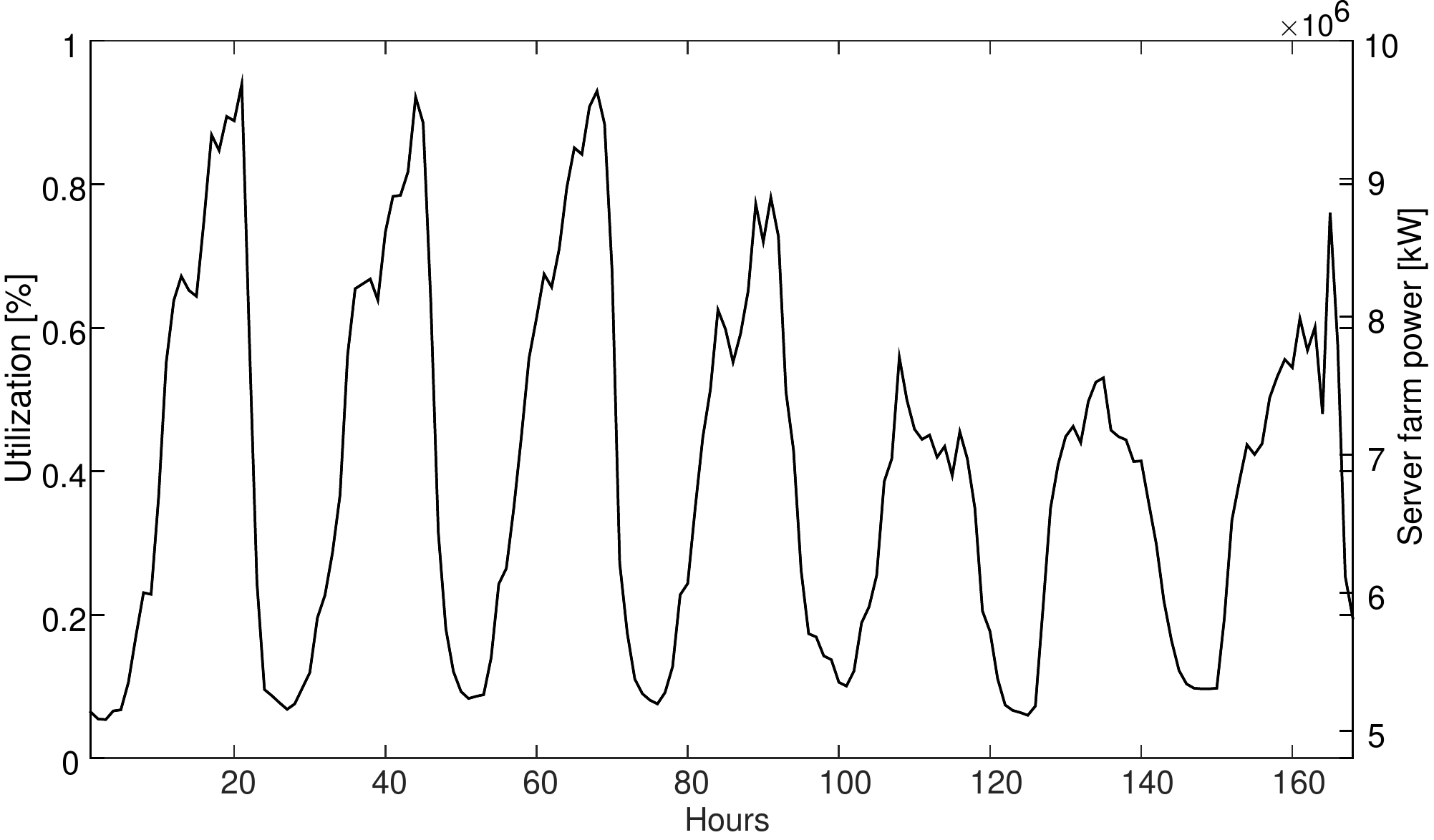}%
	%\subfloat[Case II]{\includegraphics[width=2.5in]{subfigcase2}%
	%\label{fig_second_case}}}
	\caption{Utilisation and power consumption profile of the servers in accordance with the utilisation for one week.}
	\label{utilisationAndServerPower}
\end{figure}

\begin{figure}[ht]
	\centering
	\includegraphics[width=3in]{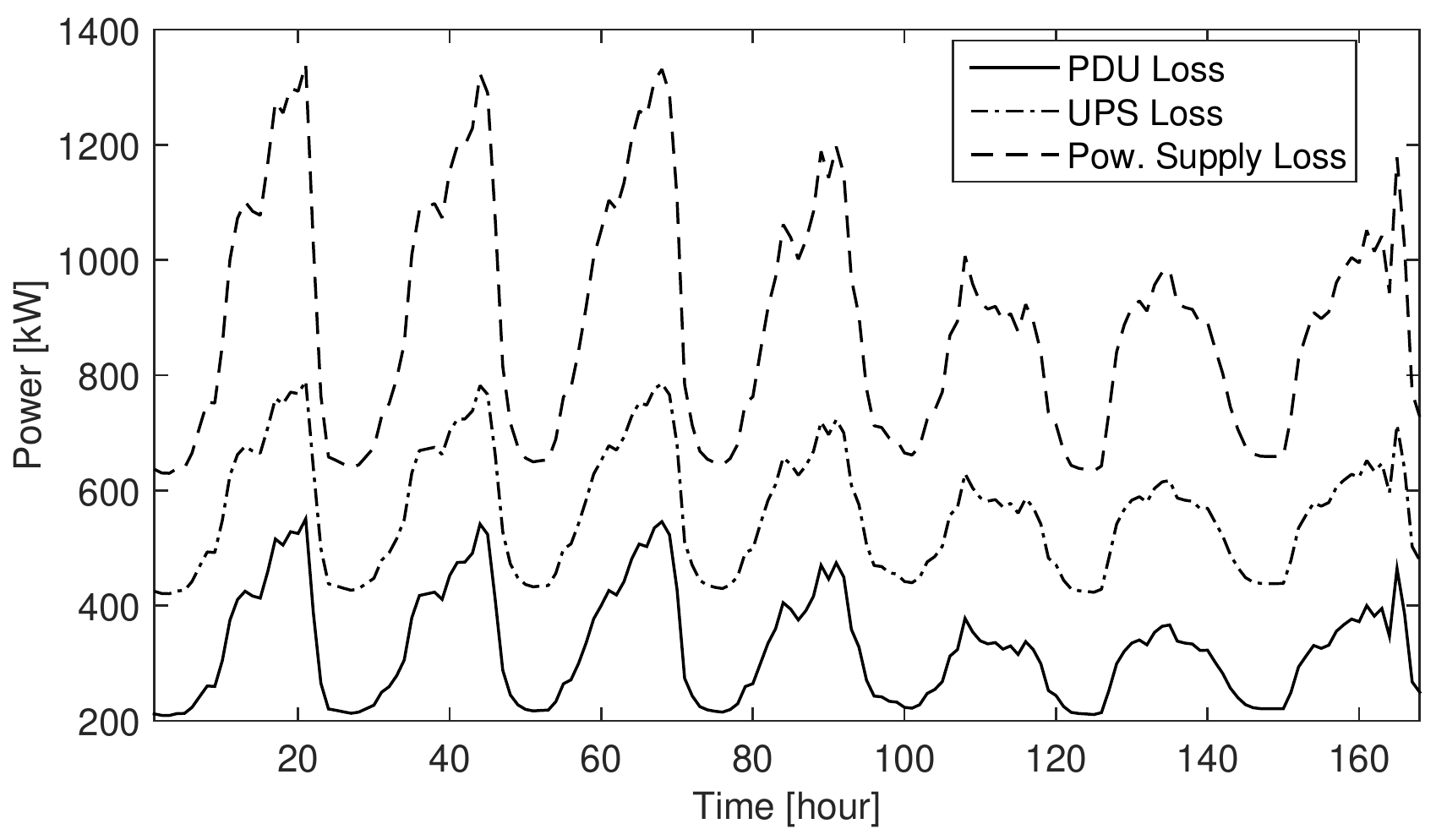}%
	%\subfloat[Case II]{\includegraphics[width=2.5in]{subfigcase2}%
	%\label{fig_second_case}}}
	\caption{Power supply loss profile of the data centre for one week.}
	\label{powerSupLoss}
\end{figure}

As a case study for modelling, we assume a data centre with a structure shown in Fig. \ref{powerFlow} and a server farm consisting of 40,000 computer servers. The cooling system of the data centre consists of a chiller plant and a CRAH unit which was chosen due to its popularity. The calculations of the physical size of the computer room and server racks are out of the scope of this research. Each server's power consumption can vary between 120 W (at idle mode) and 250 W which is within the range of Dell PowerEdge servers \cite{fitch2012dell}. As explained in Section \ref{serverFarmSection}, the key parameter in the power load model of the server farm and the whole data centre is the utilisation of servers. The hourly utilisation profile of the data centre workload for one week is assumed to be as shown in Fig. \ref{utilisationAndServerPower} (left axis) \cite{macias2014sla}. It is also assumed that the server clusters are homogeneous and load balancing is perfect. Fig. \ref{utilisationAndServerPower} (right axis) demonstrates the power consumption profile of the server farm ($\mathcal{P}_{sf}$) based on the utilisation profile and explanations provided in Section \ref{serverFarmSection}. The power loss of power supply equipment is also obtained using the equations presented in Section \ref{powerSupplySection} and shown in Fig. \ref{powerSupLoss} which consists of PDU and UPS losses as well as total loss for power supply equipment.

\vspace{-.2in}

\begin{figure}[ht]
	\centering
	\includegraphics[width=3in]{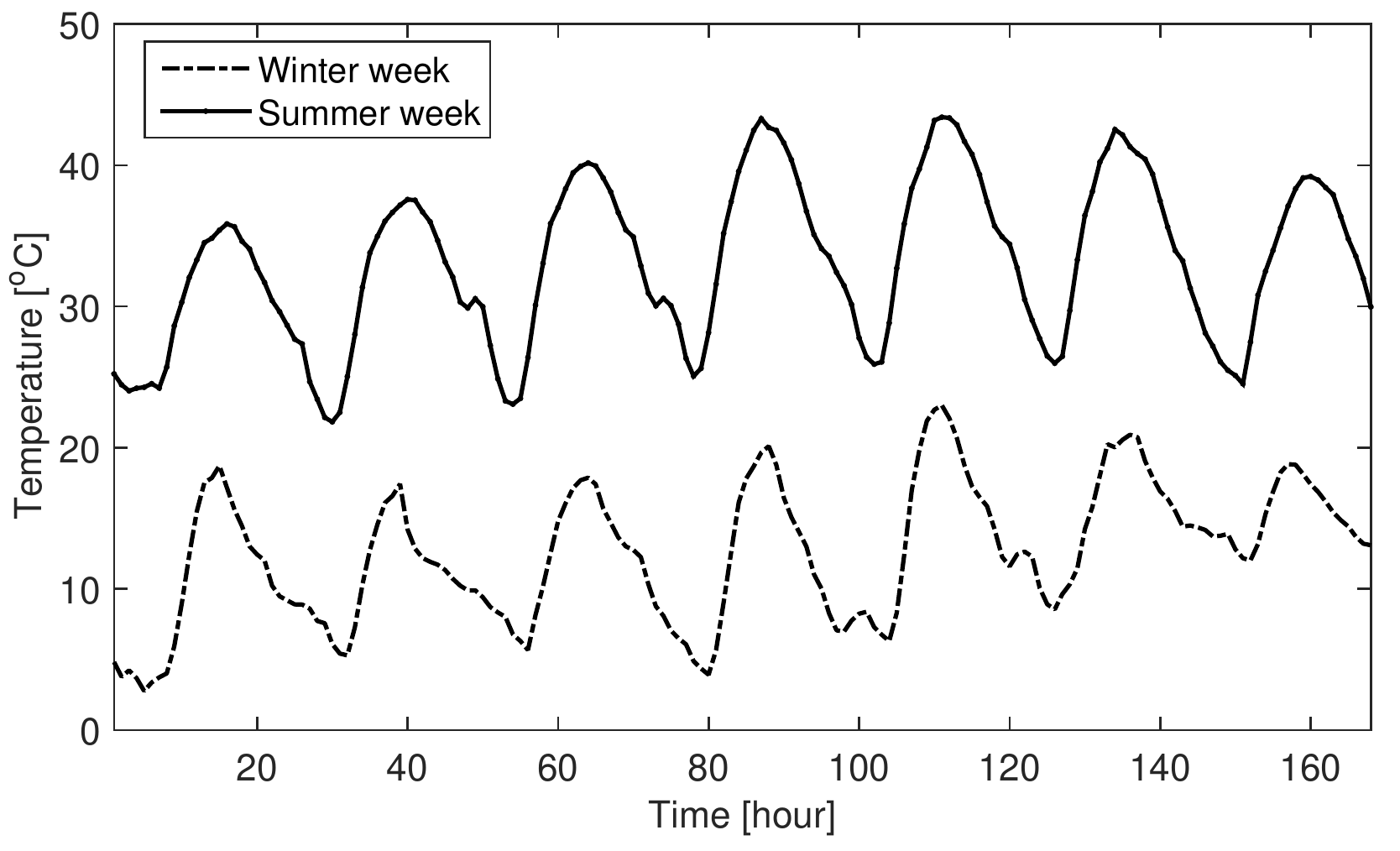}%
	%\subfloat[Case II]{\includegraphics[width=2.5in]{subfigcase2}%
	%\label{fig_second_case}}}
	\caption{Hourly outdoor ambient temperature of data centre for one week.}
	\label{temperatureProf}
\end{figure}

In order to calculate and plot the power consumption profile of the cooling system, the hourly outdoor temperature is required, as discussed in Sections \ref{chillerSection} and \ref{environImpactSection}. The weather information used in this research is obtained from the National Renewable Energy Laboratory (NREL) website \cite{NREL}. Fig. \ref{temperatureProf} demonstrates the hourly temperature profile for the first week of June 2016 (summer week) and December 2016 (winter week).

\begin{figure}[ht]
	\centering
	\includegraphics[width=3in]{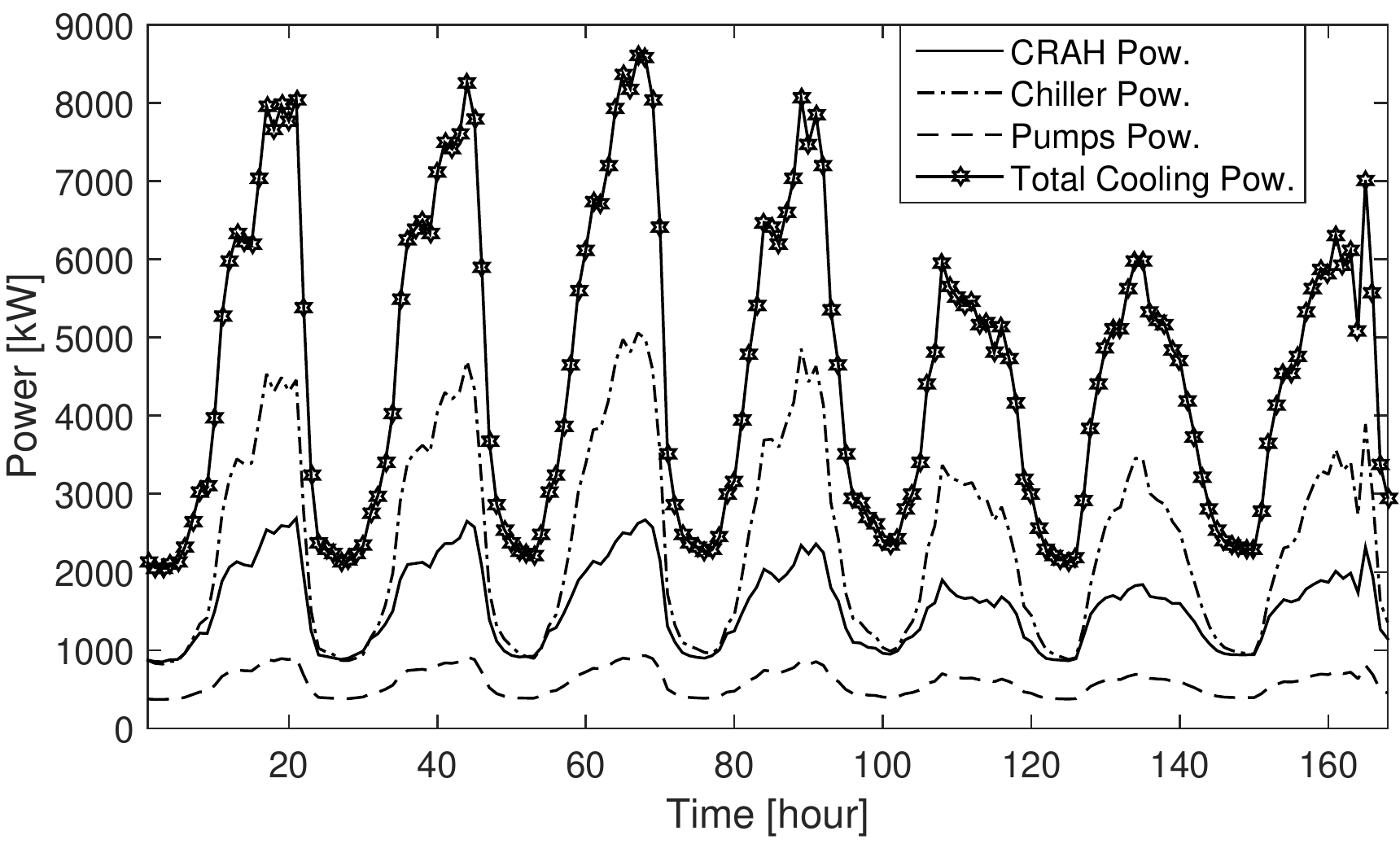}%
	%\subfloat[Case II]{\includegraphics[width=2.5in]{subfigcase2}%
	%\label{fig_second_case}}}
	\caption{Cooling system power consumption for one week.}
	\label{coolingPowerProf}
\end{figure}

The power consumption profile for the cooling system consisting of chiller plant, CRAH unit and pumps during the summer week are shown in Fig. \ref{coolingPowerProf}. The figure reflects that whenever the utilisation is low, the power consumption of the CRAH unit and the chiller plant are about the same; however, at the peak of utilisation, the power consumption of the chiller plants increases to almost twice that of the CRAH unit. Overall, the increase in power consumption of the cooling system when the utilisation changes from the minimum to its peak can reach 325\%, while the cooling system never consumes less than about 2000 kWh. It can be observed that at low server utilisation, the majority of power consumption regards to the server farm while at peak utilisation, the server farm consumes less than 50\% of the total power consumed.

\vspace{-0.1in}

\begin{figure}[ht]
	\centering
	\includegraphics[width=3in]{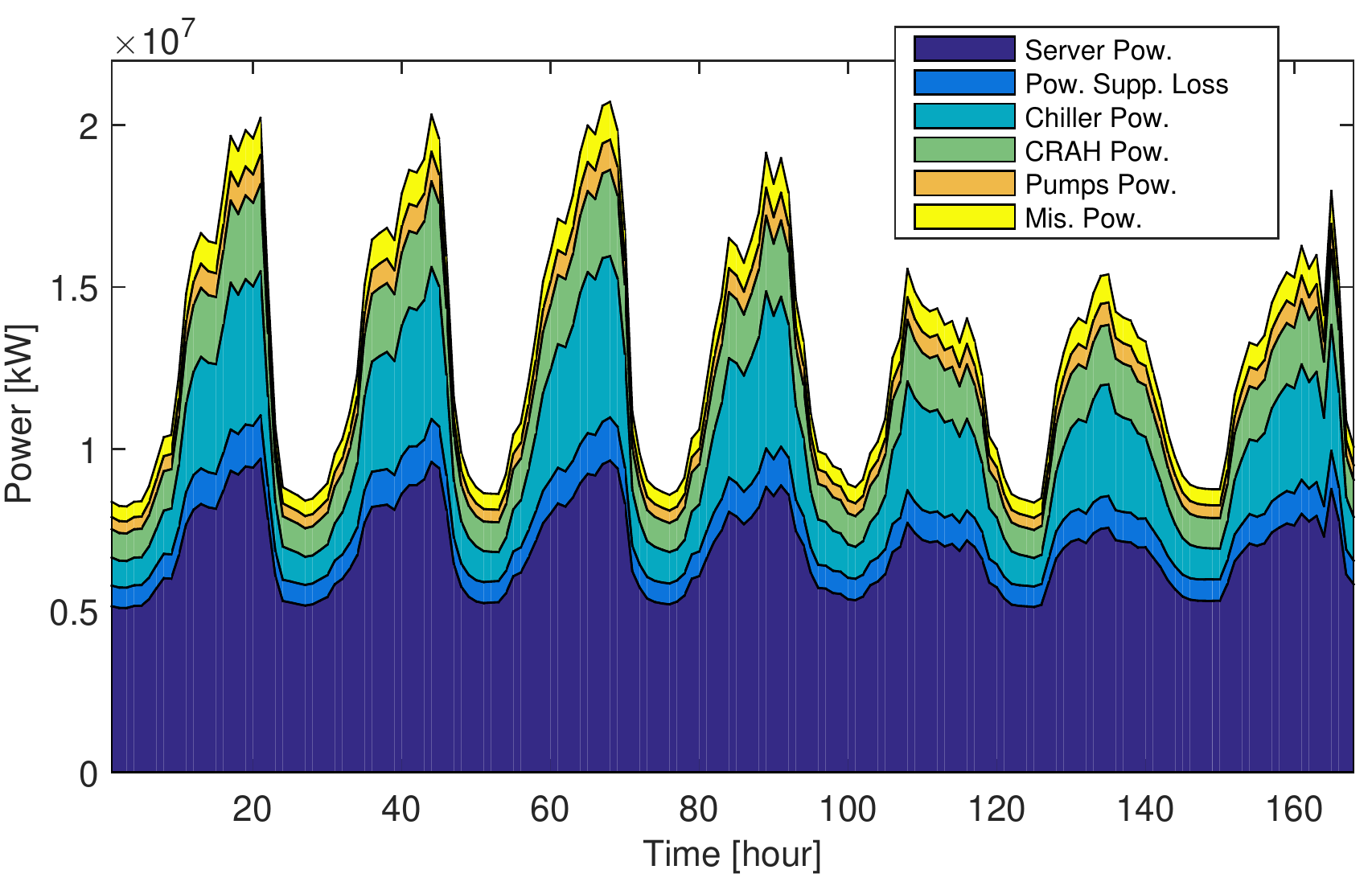}%
	%\subfloat[Case II]{\includegraphics[width=2.5in]{subfigcase2}%
	%\label{fig_second_case}}}
	\caption{Total power consumption profile and its breakdown for one week.}
	\label{totalPowerProf}
\end{figure}

In order to be able to visually compare the hourly power consumption of different components, the area plot of the data centre power consumption is shown in Fig. \ref{totalPowerProf}.

\vspace{-0.1in}

\begin{figure}[ht]
	\centering
	\includegraphics[width=3in]{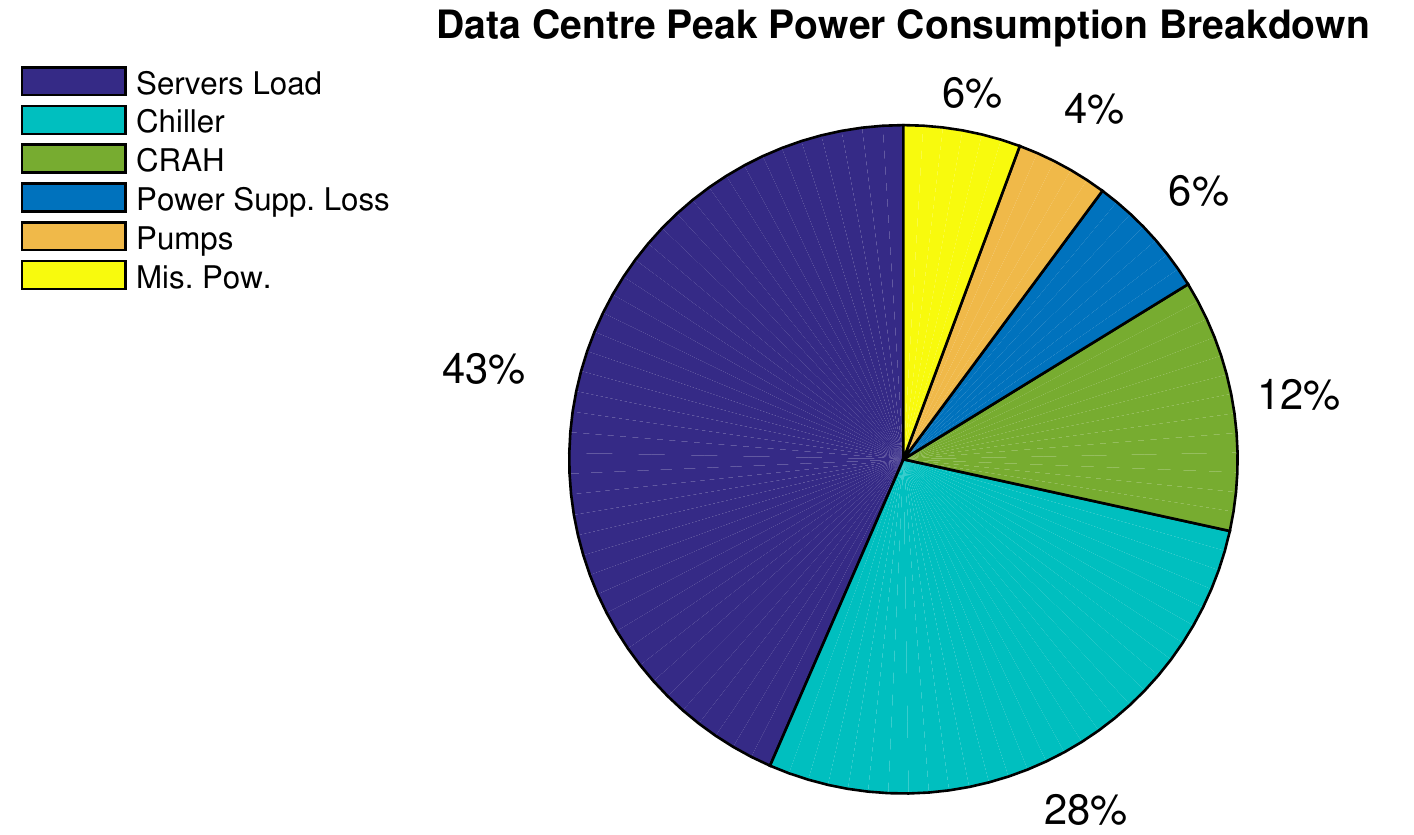}%
	%\subfloat[Case II]{\includegraphics[width=2.5in]{subfigcase2}%
	%\label{fig_second_case}}}
	\caption{Total power consumption breakdown at peak power consumption for the data centre.}
	\label{peakPowerBreak}
\end{figure}

Analysing the break-down of total power at the peak consumption is very important for the green data centres which are designed to curtail some power load whenever the power system cannot handle the load demand or the supplying microgrid faces shortfall in renewable energy generation. According to the pie chart shown in Fig. \ref{peakPowerBreak}, a server farm consumes the highest power at the peak total consumption with 43\% followed by chiller plant and CRAH unit with 28\% and 12\% respectively. Although the server farm consumes the most energy of all components in a data centre, the chiller plant, CRAH unit and pumps together consume 44\% of the total which means that the cooling system in general consumes more energy than the server farm at the peak consumption. 

\subsection{Analysing the green data centre load curtailment}

\begin{figure}[ht]
	\centering
	\includegraphics[width=3in]{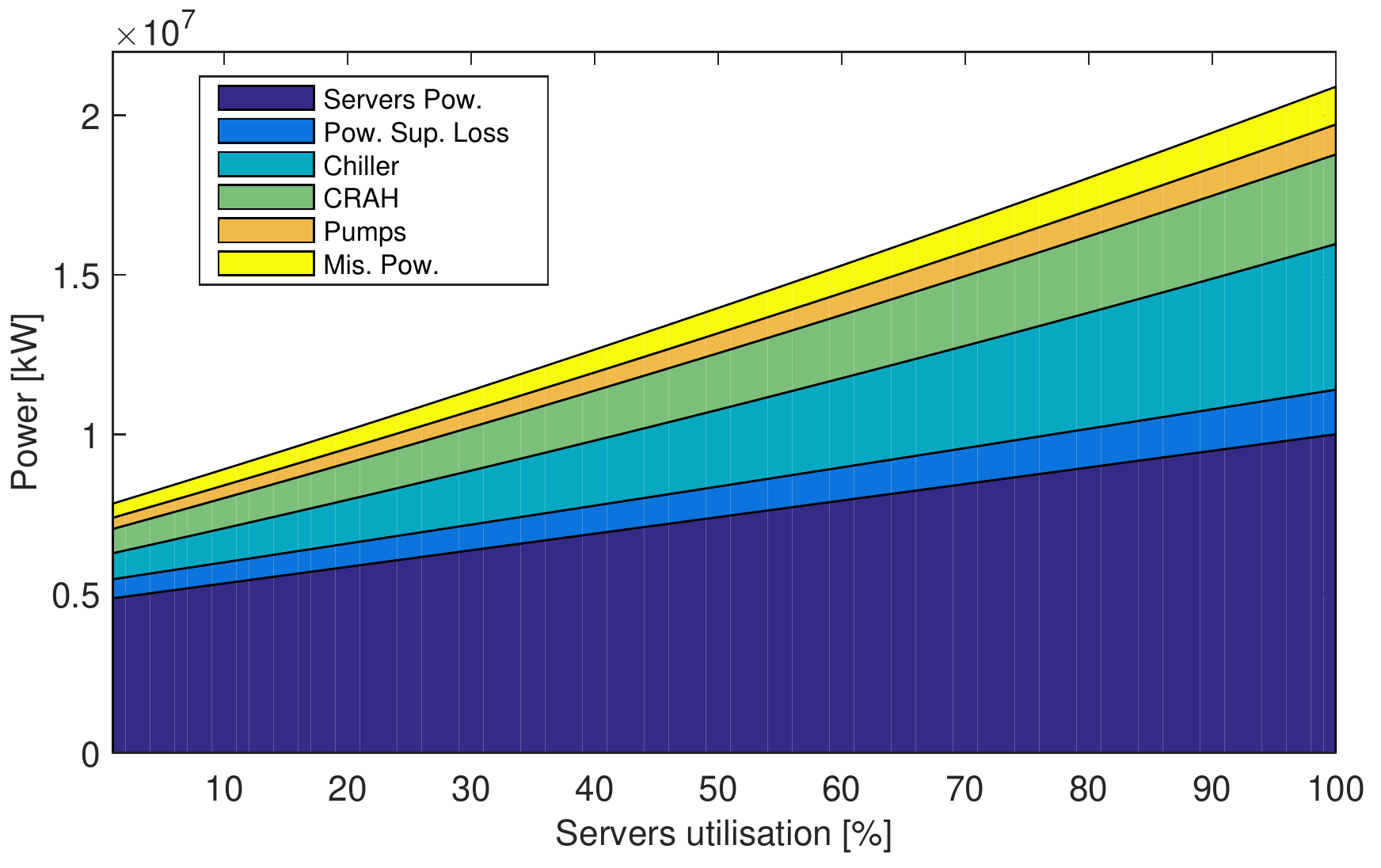}%
	%\subfloat[Case II]{\includegraphics[width=2.5in]{subfigcase2}%
	%\label{fig_second_case}}}
	\caption{Total power consumption breakdown based on the utilisation of servers for the data centre at T=$30^oC$.}
	\label{totalPowerVsUtil}
\end{figure}

One of the potential applications of the proposed model is in green data centre load curtailment of smart grids \cite{evangelopoulos2016optimal}. As explained and derived in Sections \ref{modelSection} and \ref{environImpactSection}, we can only affect the power consumption of the cooling system either by changing the server farm power or outdoor temperature from which the latter one seems impractical. As the power consumption of the server farm ($\mathcal{P}_{sf}$) entirely depends on the utilisation of the servers (Section \ref{serverFarmSection}), it is feasible to alter the servers utilisation if we need to decrease the total power consumption of a green data centre. But how does the utilisation influence the total power consumption of a data centre at a certain outdoor temperature? The answer to this question is in the area plot shown in Fig. \ref{totalPowerVsUtil}. It can be observed that by decreasing the server utilisation from 100\% to about zero, the data centre power decreases from 2.09$\times$10$^7$ to 7.8$\times$10$^6$ which shows a 55\% decrease. The plot shown in Fig. \ref{totalPowerVsUtil} is useful for determining the server utilisation respective to a certain amount of power load curtailment. As an example, if the data centre is working at 90\% utilisation and it needs to reduce the total power by 5000 MW, the server utilisation has to reduce to 54\%.  
 
\vspace{-0.2in}

 \vspace{-0.1in}

\subsection{Analysing the impact of outdoor temperature}
 
\begin{figure}[ht]
  	\centering
  	\includegraphics[width=3in]{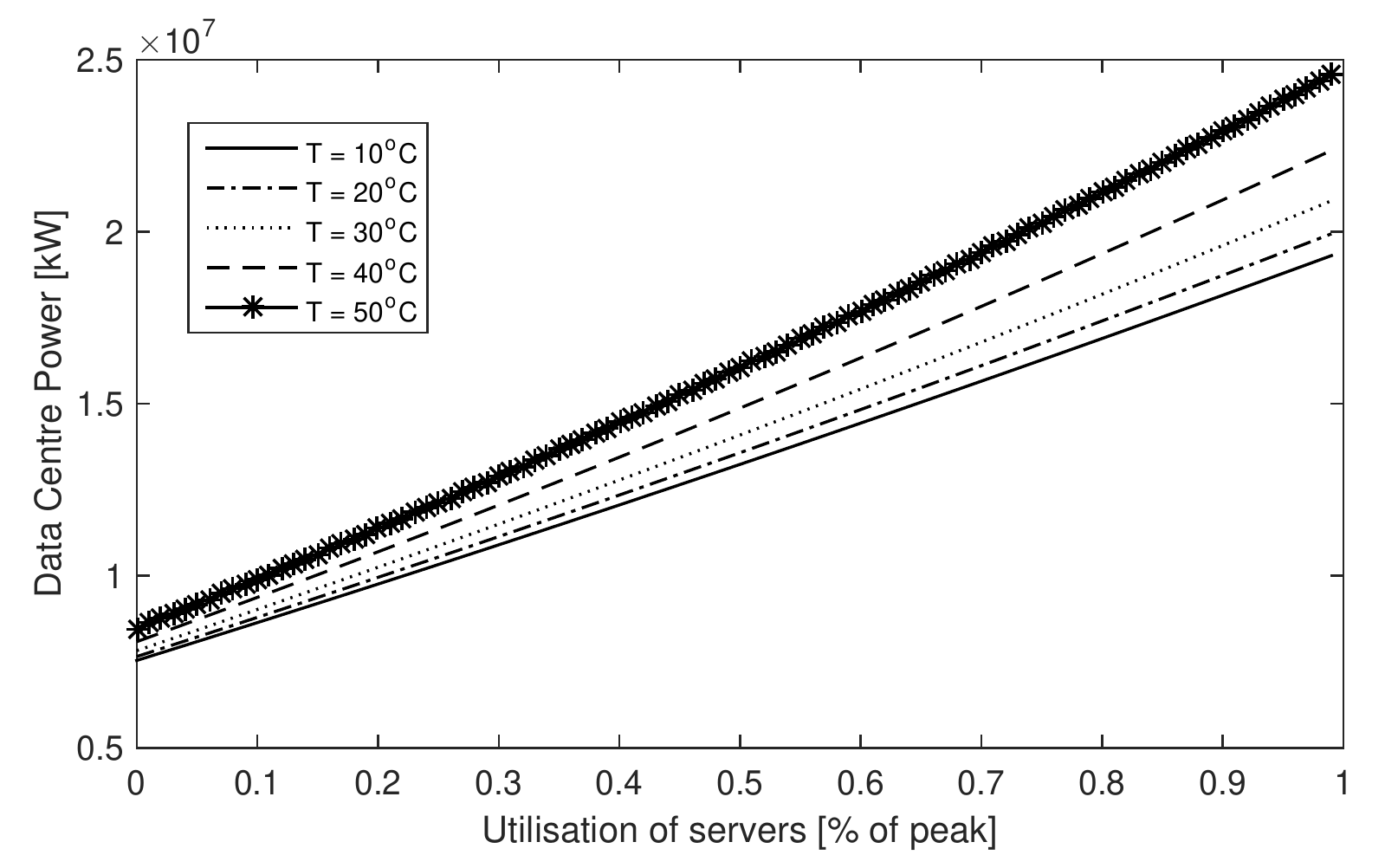}%
  	%\subfloat[Case II]{\includegraphics[width=2.5in]{subfigcase2}%
  	%\label{fig_second_case}}}
  	\caption{Total power consumption based on the utilisation of servers for different outdoor temperatures.}
  	\label{totalPowerVsUtilTemps}
  \end{figure}

The outdoor temperature of the data centre has a direct impact on the power consumption of the cooling system, and therefore a considerable impact on the total power consumption. If we run the simulation based on the temperature profile of a week in winter shown in Fig. \ref{temperatureProf}, it is clear that the contribution of the chiller plant to the data centre energy consumption at peak decreases to 16\% which is a change from 4.026$\times$10$^8$ to 2.5884$\times$10$^8$ kWh compared with a summer week. Also, the total energy consumption of the data centre in a week in summer is 7.88\% bigger than that of a week in winter, totalling 2.0306$\times$10$^9$ kWh. The total power consumption profile of the data centre versus servers utilisation is plotted for various outdoor temperatures and shown in Fig. \ref{totalPowerVsUtilTemps}. The usage of this plot is the same as Fig. \ref{totalPowerVsUtil} while it provides the outdoor temperature option as well.

\vspace{-0.1in}
 
  \subsection{Analysing the annual energy consumption}
  
\begin{figure}[ht]
  	\centering
  	\includegraphics[width=3in]{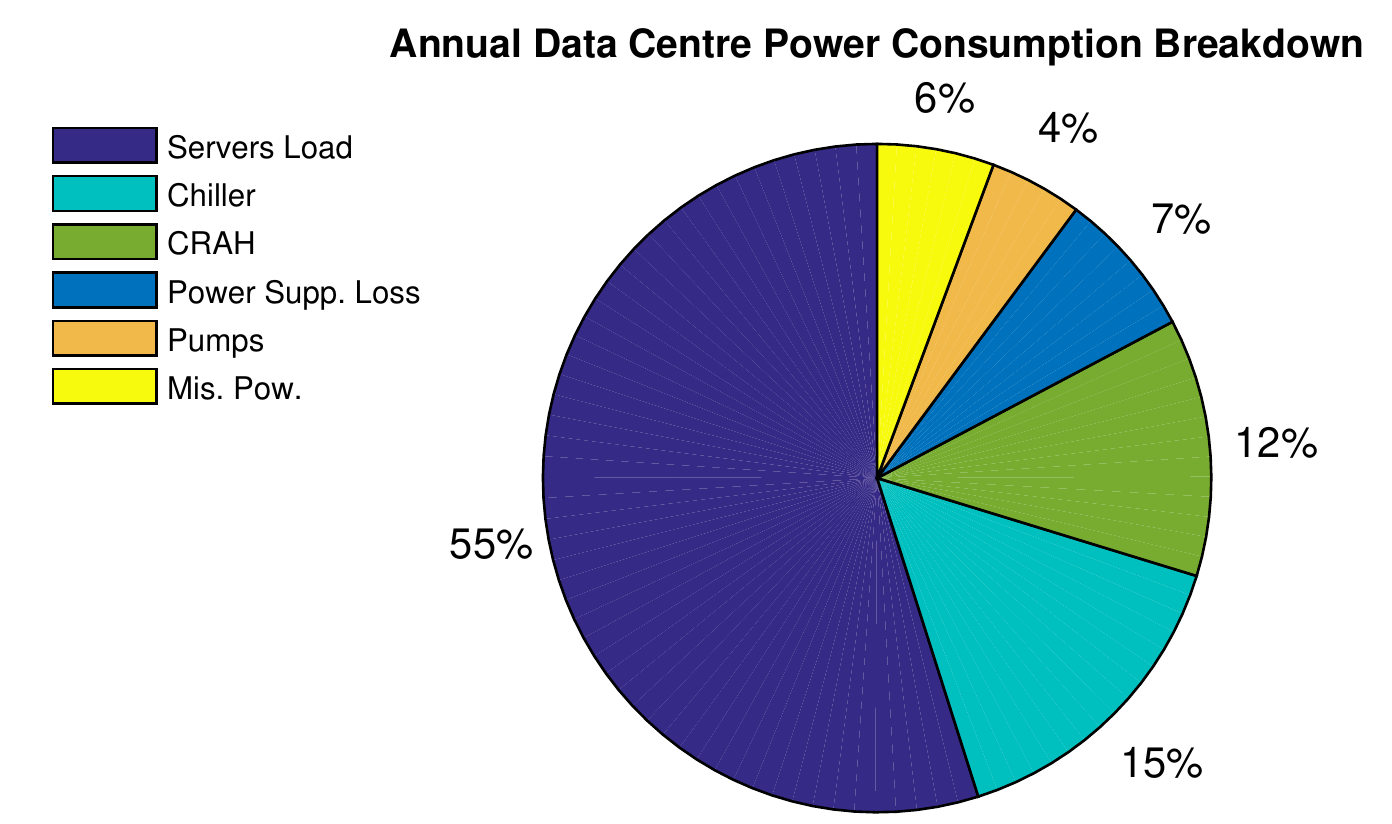}%
  	%\subfloat[Case II]{\includegraphics[width=2.5in]{subfigcase2}%
  	%\label{fig_second_case}}}
  	\caption{Annual power consumption breakdown for the data centre.}
  	\label{annualPowerBreak}
  \end{figure}

  In order to analyse the annual energy consumption profile of the data centre, an hourly utilisation profile is generated using the pattern of the weekly utilisation profile. Also, the hourly temperature profile for 2016 is obtained from NREL website and assumed as outdoor temperature of the data centre \cite{NREL}. The breakdown of the annual energy consumption is shown as a pie chart in Fig. \ref{annualPowerBreak}. In can be observed that 55\% of the annual energy is consumed by the server farm itself. The chiller plant consumes the second highest portion of energy by 15\% followed by 12\% for the CRAH unit. The cooling system consumes 31\% of the total annual energy of the data centre which is 1.0937$\times$10$^{11}$ kWh.
The proposed model has made it possible to model the annual consumption which is essential for strategic planning as the impacts of different weather conditions are already included in the figures obtained.

\vspace{-0.1in}
\subsection{Analysing the impact of architecture on energy consumption}

\begin{figure}[ht]
	\centering
	\includegraphics[width=3in]{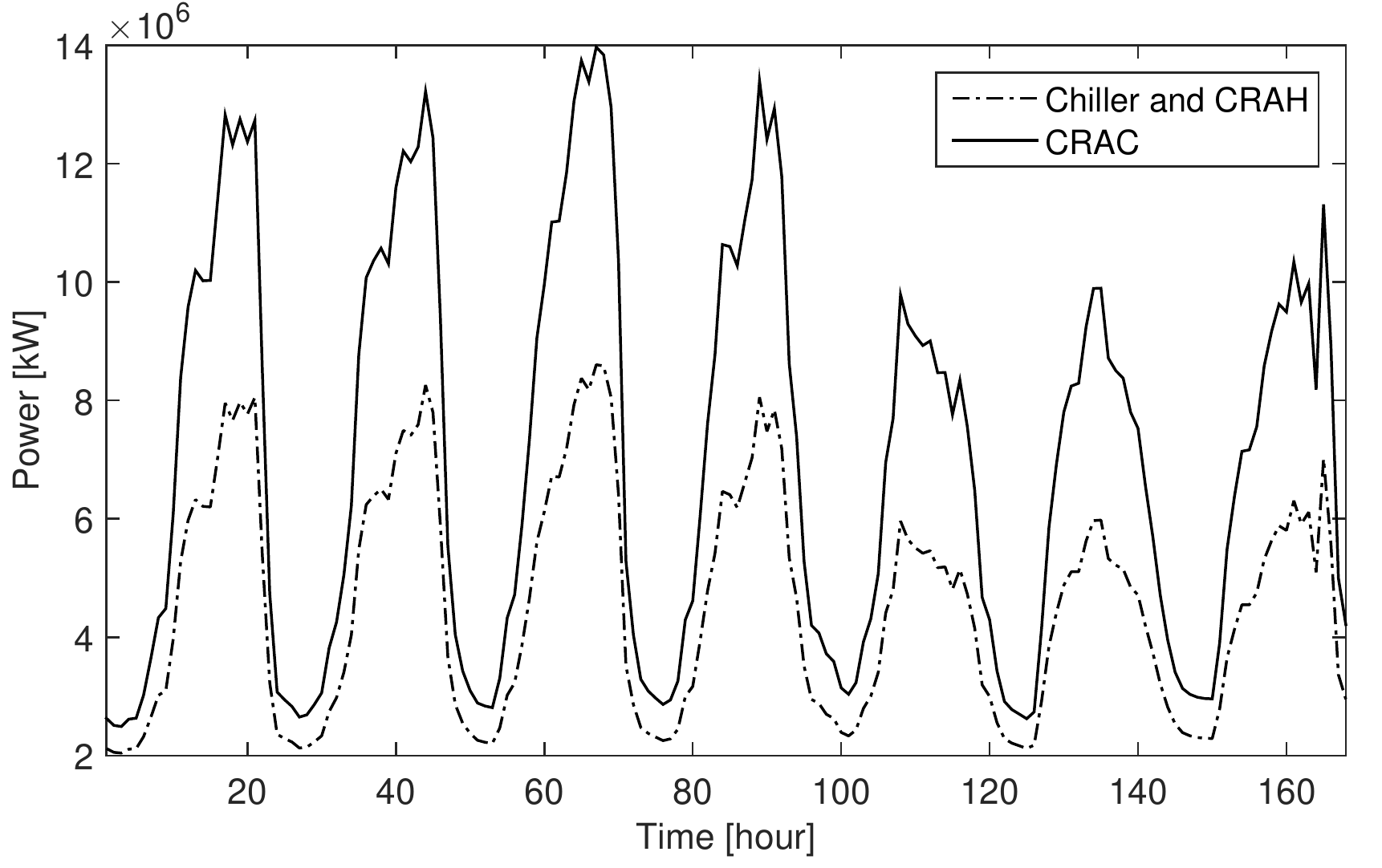}%
	%\subfloat[Case II]{\includegraphics[width=2.5in]{subfigcase2}%
	%\label{fig_second_case}}}
	\caption{Hourly power consumption profile for different cooling systems during summer week.}
	\label{CRACVsChiller}
\end{figure}

%\hl{Since the server farm has a fixed architecture, the only component which can have noticeable impacts on the total energy consumption of the data centre by its variations is the cooling system. This is due to its high contribution to the energy consumption.}
For any given server farm size, only different structures of the cooling system can have noticeable impacts on the total energy consumption of the data centre, due to its high contribution to the energy consumption.
If we change the architecture of the cooling system for the data centre under analysis by replacing the chiller plant and CRAH unit with a CRAC unit the hourly power consumption profile for summer week will be as shown in Fig. \ref{CRACVsChiller}. Installing a CRAC unit increases the power consumption of the cooling system by 44.22\% and 55.52\% for summer and winter week respectively. Also it increases the annual energy consumption of the cooling system by 49.38\% which yields its annual contribution in energy consumption to 42\%.

\vspace{-0.2in}
\section{Conclusion}
%\blindtext
In this paper, a complete model for modular simulation of the modular design of a data centre is presented. The model consists of a detailed analysis of the power consumption of each component as well as the interactions between the components. Unlike existing research in the literature, the present modular simulation model can take different design structures of data centre into account and provide us with hourly power consumption profiles for each component. 
Based on the analysis performed, the server farm is the most significant energy consumer of a data centre with more than 50\% contribution annually which is followed by the cooling system. The server farm and cooling system together consume about 90\% of the data centre energy consumption which denotes their roles in energy management plans of a data centre. The ambient temperature of the data centre considerably affects the power consumption of cooling system. The simulation results show that the energy consumption of the chiller plant can increase by about 8\% if the ambient temperature of a week in summer is given instead of winter. This can contribute about 15\% of the whole data centre energy consumption. In contrast, the impact of outside data centre humidity on the energy consumption of a data centre is shown to be negligible. Also, it is obtained that using CRAC system instead of chiller plant and CRAH units can increase the annual energy consumption of the cooling system by about 50\%. The application of the proposed model for green data centre load curtailment is proposed and investigated as well.
%The impacts of environmental parameters such as temperature and humidity are also investigated and incorporated into the model. It is worth mentioning that the present power consumption model is independent of the main source of electricity of the data centre which can be supplied by the main grid or by renewable energies through microgrids. The flexibility, scalability, comprehensiveness and modularity of this model provides the researchers and designers with a powerful tool for energy analysis, management and planning of data centres with different designs and locations.

\ifCLASSOPTIONcaptionsoff
  \newpage
\fi



\bibliographystyle{IEEEtran}
\bibliography{biblio}

%	Biography text here.
%\end{IEEEbiography}

%	Biography text here.

%\newpage

%	Biography text here.


%\vfill


\end{document}